\documentclass[12pt,nofootinbib]{revtex4}
\begin{document}
\title{On the Meaning of the Principle of General Covariance}
\author{Alberto Chamorro}
\affiliation{Department of Theoretical Physics,
University of the Basque Country,
48080 Bilbao, Spain}
\email{alberto.chamorro@ehu.es}

\maketitle

\section*{Abstract}
We present a definite formulation of the Principle of General Covariance (GCP) as a Principle of General Relativity with physical content and thus susceptible of verification or contradiction. To that end it is useful to introduce a kind of coordinates, that we call quasi-Minkowskian coordinates (QMC), as an empirical extension of the Minkowskian coordinates employed by the inertial observers in flat space-time to general observers in the curved situations in presence of gravitation. The QMC are operationally defined by some of the operational protocols through which the inertial observers determine their Minkowskian coordinates and may be mathematically characterized in a neighbourhood of the world-line of the corresponding observer. It is taken care of the fact that the set of all the operational protocols which are equivalent to measure a quantity in flat space-time split into inequivalent subsets of operational prescriptions under the presence of a gravitational field or when the observer is not inertial. We deal with the Hole Argument by resorting to the idea of the QMC and show how it is the metric field that supplies the physical meaning of coordinates and individuates point-events in regions of space-time where no other fields exist. Because of that the GCP has also value as a guiding principle supporting Einstein's appreciation of its heuristic worth in his reply to Kretschmann in 1918.

\section{Introduction}

Since first formulated nine decades ago the question of the meaning of the GCP has been a subject of polemic and confusion. Thus Kretschmann~\cite{kr} in 1917 claimed the GCP to be devoid of physical content and that given enough mathematical ingenuity any theory could be set in a general covariant form. Einstein~\cite{ei} begrudgingly accepted the objection stating however the heuristic value the GCP had in searching for a good theory and that that was a reason to prefer General Relativity to Newtonian gravitation which -in his opinion- would only be awkwardly casted into generally covariant form. Einstein was soon proved wrong as Cartan~\cite{ca} in 1923 and Friedrichs~\cite{fr} in 1927 found serviceable generally covariant formulations of Newtonian gravitation theory. See also Misner \emph{et al}(1973, ch 12)~\cite{mi}. In his excellent book Fock~\cite{fo} makes interesting and critical remarks about the term ``general relativity'' adopted by Einstein to name his theory of gravitation and the connection of the term with general covariance that, in his view, is merely a logical requirement that is always satisfiable. Fock rightly points out that though Einstein had agreed with Kretschman's objection as to the physical vacuity of the GCP his agreement was rather formal, because actually to the end of his life Einstein related the requirement of general covariance to the idea of some kind of ``general relativity'' and with the equivalence of all frames of reference. The subject has subsequently been addressed in several ways, for example,  by Anderson~\cite{an}(1967), Stachel~\cite{st}(1980, 1986, 2002), Norton~\cite{no}(1993) and Ellis and Matravers ~\cite{el}(1995). All these works while attempting to clarify the formulation and meaning of the GCP in our opinion fail to give it a specific expression susceptible of physical verification or contradiction. And certainly whatever the claim about the physical content of the GCP might be that should be subject to experimental test to be confirmed or refuted. In fact, though not directly dealing with the GCP but acknowledging that it has a conceptual content far deeper than the simple invariance under arbitrary changes of coordinates, Lusanna and Pauri~\cite{lu} (2006) go a significant step beyond previous authors considering the physical individuation of space-time points by experimental procedures and revise the Hole Argument. We fully subscribe to their contention that ``the gauge freedom of general relativity is unavoidably entangled with the definition-constitution of the very \emph{stage}, space-time, where the \emph{play} of physics is enacted'' and lend further support to that idea by discussing the Hole Argument using operationally defined coordinates introduced in our formulation of the GCP. Ellis and Matravers~\cite{el} point out how physicists and astrophysicists in fact almost always use preferred coordinate systems not merely to simplify the calculations but also to help define quantities of physical interest, and that this suggests that we should reconsider and perhaps refine the dogma of general covariance. In that spirit we present in this contribution a proposal for the GCP and show that it has two meanings: a predictive one as a principle of general relativity, that in principle may be falsifiable by resorting to experience, and a heuristic one as a guiding principle to extend the theory and probe the nature of space-time.\\

The plan of the paper is as follows:
\begin{itemize}
\item In Section \ref{sec:GCP} we define a principle of general relativity and take it as the GCP. 
\item In Section \ref{sec:QMC} we construct a family of coordinates which are useful to endow with physical meaning the GCP.
\item Section \ref{sec:hole} presents a treatment of the Hole Argument with the help of the mentioned family of coordinates and considers the implications of the GCP on the meaning of coordinates and point-events in space-time.
\item In Section \ref{sec:remarks} some other consequences of our formulation of the GCP are indicated and finally our conclusions regarding the meaning of the GCP are stated. 
\end{itemize}

\section{The GCP as a Principle of General Relativity}\label{sec:GCP}
\subsection{Principles of Restricted Relativity (RRP) and the Principle of Special Relativity (SRP)}

We will first formulate a principle of restricted relativity (RRP) with
respect to a group of isometric diffeomorphisms of space-time: Let us
have a generic space-time $({\bf M}, {\bf g})$, ${\bf M}$ and ${\bf g}$
respectively denoting the manifold and the metric. Let $x^{\lambda},
\lambda=0,1,2,3$, be the coordinates corresponding to some neighbourhood
$N$ of an  arbitrary point $P$ of {\bf M} and let $F(P)$ be a physical
quantity -that might have several components- defined at $P$ and possibly on $N$. Let us denote by  $F(P,x)$ the values at $P$ of $F(P)$ in the coordinates $x^{\lambda}$. Let $Q[F(P,x)]$ stand for
the set of all the different operational protocols -but equivalent in
the sense that they yield the same values- that may be used to determine
$F(P,x)$ 
\footnote{For instance, if {\bf E} is the electric field, $Q[{\bf E}(P,x)]$ would be the set of all possible measurement methods to determine the components of {\bf E} at $P$ in the coordinates $x^{\lambda}$}.
The physical
meaning of $F(P)$ is clearly given by $Q[F(P,x)]$ for any given $x^{\lambda}$. In fact since the values $F(P,x')$ of $F(P)$ in another coordinates $x'^{\lambda}$ can be obtained from the values $F(P,x)$ and the corresponding coordinate transformation rules for $F(P)$ it is enough to consider $Q[F(P,x)]$ for some $x^{\lambda}$ to have completely defined the physical meaning of $F(P)$. If there is a group
$\mathcal{L}$ of isometries of ${\bf g}$ we shall say that a RRP with
respect to $\mathcal{L}$ exists if the following two conditions hold:
\begin{itemize}
\item[(a)] The equations describing the behaviour of the physical
quantities are form-invariant under all the transformations induced by
the elements of $\mathcal{L}$.

\item[(b)] One has that if $\Lambda\in \mathcal{L}$ and $\Lambda:x
\longrightarrow x'$, $\Lambda:F(P,x)\longrightarrow F'(P,x')$, then
$Q[F(P,x)]=Q[F'(P,x')]\ \forall\ \Lambda\in \mathcal{L}~ \textrm{and}
~\forall ~F(P).$ Note in particular that
$Q[x^{\lambda}]=Q[x'^{\lambda}]$, that is, the primed and unprimed coordinates should also be determined by identical sets of operational protocols.
\footnote{Wald calls principle of special covariance to a RRP ~\cite{wa}.}.
\end{itemize}
A RRP is a symmetry principle with a clear physical meaning contained in
the above two conditions. If ${\bf g}$ has no isometry there is no RRP
in the defined sense. The principle of special relativity is a RRP with
respect to the \emph{proper} Poincar\'e group of transformations%
\footnote{The improper elements of the full Poincar\'e group should be
excluded on account of the existence of phenomena that violate parity
and or time reversal symmetry.}. Let $({\bf M}, {\bf \eta})$ be flat
space-time of special relativity and let $\mathcal{I}$ stand for the set
of all the inertial observers in it. Each of these observers is supposed
to be located at the origin of a non-rotating Cartesian frame at rest in
his or her proper reference frame with the help of which he or she
assigns the spatial coordinates to events. The time coordinate of any
event is assigned by each observer by the standard criteria of special
relativity. The resulting coordinates are the Minkowskian coordinates in
which the metric tensor is $\eta_{\mu\nu}=diag(+1,+1,+1,-1).$ Let $O$
and $O'$ be any two observers of $\mathcal{I}$ and $x$ and $x'$ their
respective Minkowskian coordinate systems, related by a transformation of the proper Poincar\'e group. Then the SRP may be defined
more specifically as follows: 
\begin{itemize}
\item[(a)] The equations describing the behaviour of the physical quantities have the same form for $O$ and $O'$ when expressed in terms of $x$ and $x'$. 
\item[(b)] If $F(P,x)$ and $F'(P',x')$ stand for the same physical quantity respectively measured by $O$ and $O'$ at points $P$ and $P'$ using Minkowskian coordinates $x$ and $x'$, $Q[F(P,x)]=Q[F'(P',x')]\ \forall ~F$ and $\forall ~P$ and $P'$ where $F$ might be defined.
\end{itemize}

It is quite easy to see that the SRP so defined fulfills the conditions required above to have a RRP.

\subsection{Principle of General Relativity (GRP)}\label{ssec:GRP}

We will introduce the GRP as a generalization in two directions of the SRP as previously defined. Both restrictions, that of the existence of an isometry and that of refering  to measurements performed only by inertial observers will be relinquished. The set $\mathcal{I}$ will be enlarged to the class of observers, $\mathcal{K}$, having world-lines as differentiable as it may be needed for the subsequent developments. The Minkowskian coordinates employed by the inertial observers of $\mathcal{I}$ will be generalized to coordinates used by the observers of $\mathcal{K}$ operationally defined by some of the operational protocols through which the observers of $\mathcal{I}$ determine their Minkowskian coordinates. These more general coordinates will be called quasi-Minkowskian (QMC). In other words, if $\tilde{x}^{\lambda}$ and $x^{\lambda}$ respectively are quasi-Minkowskian and Minkowskian, it should hold that $\tilde{Q}[\tilde{x}^{\lambda}] \subseteq Q[x^{\lambda}]$, where the left-hand term denotes the different procedures of measurement of $\tilde{x}^{\lambda}$ by the observer of $\mathcal{K}$ in question whereas the right-hand term stands for the analogous set for the $x^{\lambda}$'s and any observer of $\mathcal{I}$. The latter relation follows from the expectation that the set $Q[x^{\lambda}]$ should split into inequivalent subsets of operational prescriptions under the presence of a gravitational field or when the observer is not inertial~\cite{de}. Let us then first characterize mathematically the QMC: 

Let $O$ be an observer in a generic space-time $({\bf M}, {\bf g})$ following a world-line $C$ given by its equations $x^{\lambda}=f^{\lambda}(\tau)$, where $\tau$ is $O$'s proper time.
$O$'s four-velocity is $u^{\lambda}=\displaystyle\frac{dx^{\lambda}}{d\tau}=\dot f^{\lambda}(\tau)$, $u^{\lambda}u_{\lambda}=-c^{2}$. 
$O$ transports an orthonormal tetrad $\mathbf{e}_{(\nu)}$ along his world-line whose components verify
\begin{equation}e^{\lambda}_{(\nu)}e_{(\mu)\lambda}=\eta_{\nu\mu}~~,~~~e^{\mu}_{(0)}=\frac{u^{\mu}}{c}~.\end{equation}
The most general smooth transportation law of the tetrad that conserves these conditions is given by  
\begin{equation}\frac{De^{\mu}_{(\sigma)}}{d\tau}=\frac{1}{c^{2}} (u^{\mu}a^{\nu}-u^{\nu}a^{\mu})e_{(\sigma)\nu}+\frac{1}{c}\omega_{\alpha}u_{\beta}\epsilon^{\alpha\beta\mu\nu}e_{(\sigma)\nu}~,\end{equation}
where 
\begin{displaymath} a^{\nu}=\frac{Du^{\nu}}{d\tau}~,~~\epsilon_{\alpha\beta\gamma\delta}=(-g)^{\frac{1}{2}}[\alpha\beta\gamma\delta]~,~~\epsilon^{\alpha\beta\gamma\delta}=-(-g)^{-\frac{1}{2}}[\alpha\beta\gamma\delta]~,
\end{displaymath}
\begin{displaymath}
[\alpha\beta\gamma\delta]=\left\{\begin{array}{ll}
+1& \textrm{if $\alpha\beta\gamma\delta$ is an even permutation of $0123$}\\
-1 & \textrm{if $\alpha\beta\gamma\delta$ is an odd permutation of $0123$}\\
\phantom{+}0 & \textrm{if $\alpha\beta\gamma\delta$ are not all different}
\end{array} \right.
\end{displaymath}

\begin{displaymath}
g=\det\vert\vert{g_{\alpha\beta}}\vert\vert~~,\end{displaymath}  and the $\omega_\alpha~'s$ are the covariant components of a rotation vector such that $u^{\alpha}\omega_\alpha=0$.

As an example, the observer $O$ could choose a set of QMC, $\tilde{x}^{\lambda}=(c\tilde{t},\tilde{x}^{i}),~~i=1,2,3$, in the following way:
He manages to send a radar signal at his proper time  $\tau_{e}$ such that it arrives at the point-event whose QMC coordinates are to be determined just as it happens. The signal is inmediately reflected back reaching $O$ at his proper time $\tau_{r}$. $O$ defines $\tilde{x}^{0}\equiv \frac{1}{2}c(\tau_{r}+\tau_{e})$. He also sets $d\equiv \frac{1}{2}c(\tau_{r}-\tau_{e})$ for his ''distance'' to the point-event and has previously noted the direction of the emitted radar signal by recording the cosines, $\cos  \phi_{i}, i=1,2,3$ of the angles, $\phi_{i}, i=1,2,3$, that the 
signal ray respectively makes with each of the directions of the spatial triad $\mathbf{e}_{(i)}, i=1,2,3$. Then $O$ defines the remaining QMC as  $\tilde{x}^{i}\equiv d \cos  \phi_{i}, i=1,2,3$. Let $v^{\lambda}$ the components of a 4-vector along the emitted electromagnetic signal such that $v^{0}=1$. Then $\cos \phi_{i} =v^{\lambda}e_{(i)\lambda}, i=1,2,3$, and since only two of the three  $\cos \phi_{i}$ are independent it turns out that the four $\tilde{x}^{\lambda}$ can be expressed in terms of four independent invariants, $\tau_{e}, \tau_{r}, v^{\lambda}e_{(i)\lambda}, i=1,2$, 
for instance\footnote{$\tau_{e}$ and  $\tau_{r}$ are invariants once the origin for the proper time has been set which amounts to setting the origin of the time $\tilde{t}\equiv \frac{1}{c}\tilde{x}^{0}$.}. These invariants are directly associated to the tools used by $O$ to label the events but, since any other set of coordinates in a small enough neighborhood of his world-line will be functionally related to the  $\tilde{x}^{\lambda}$, any coordinates will also be so functions of the said four invariants albeit different ones. 

More generally, we will impose the following mathematical conditions to characterize the QMC coordinates, $\tilde{x}^{\lambda}=(c\tau,\tilde{x}^{i})=(c\tilde{t},\tilde{x}^{i}),~~i=1,2,3$:

\begin{enumerate}
\item $C$ is described in the $\tilde{x}^{\lambda}$ coordinates by:~~~$\tilde{x}^{i}=0~,~~~\tilde{t}=\tau$. 
\item The restriction of the metric in the $\tilde{x}^{\lambda}$ coordinates on $C$ is: $\tilde{g}_{\mu\nu}\mid_{C}=\eta_{\mu\nu}$, and\\ $\displaystyle\frac{\partial\tilde{g}_{\mu\nu}}{\partial\tilde{x}^{\lambda}}\mid_{C}=0,$ when the four-acceleration of $O$ and the four-rotation of the tetrad vanish: $\mathbf{a}=\mathbf{\omega}=0$.
\item $\mathbf{e}_{(\alpha)}= \displaystyle\frac{\partial}{\partial\tilde{x}^{\alpha}}\mid_{C}$~$\Longleftrightarrow$~ $e^{\lambda}_{(\alpha)}(\tau)=\displaystyle\frac{\partial x^{\lambda}}{\partial\tilde{x}^{\alpha}}\mid_{C}$~. 
\item The $\tilde{x}^{\lambda}$'s become the usual Minkowskian coordinates in a neighborhood of $C$ when $\mathbf{a}=\mathbf{\omega}=0$ and the curvature tensor vanishes, $R_{\alpha\beta\mu\nu}=0$, whithin that neighborhood. 
\end{enumerate}
The $\tilde{x}^{\lambda}$'s are QMC and since they are so with respect to a world-line $C$ and depend on the choice of the space triad, $\mathbf{e}_{(i)}$,
they will be denoted henceforth by QMC$C$$\mathbf{\omega}$
(quasi-Minkowskian coordinates relative to the world-line $C$ and to the
tetrad $\mathbf{e}_{(\alpha)}$ subject to the rotation
$\mathbf{\omega}$). Fermi coordinates and the coordinates introduced by
Lachi\`{e}ze-Rey~\cite{la}
\footnote{I am indebted to Llu\'is Bel
for bringing to my attention this reference.} are other particular
instances of QMC$C$$\mathbf{\omega}$'s.  

\subsubsection{Principle of General Relativity (GRP)}

Let us have a generic smooth enough space-time $({\bf M}, {\bf g})$, ${\bf M}$ and ${\bf g}$ respectively denoting the manifold and the metric. Let $O$ be any observer in that space-time belonging to the class $\mathcal{K}$ that uses any type of QMC$C$$\mathbf{\omega}$'s. If $F$ is any -generally multicomponent- physical quantity let us denote by $\tilde{F}$ the same quantity -or its components- in the QMC$C$$\mathbf{\omega}$, $\tilde{x}^{\lambda}$'s, as determined by $O$. Let us denote by $\tilde{Q}[\tilde{F}]$ the set of operational protocols that may be used by $O$ to measure $\tilde{F}$ and $Q[F]$, as usual, the analogous set of protocols used by inertial observers in flat space-time to measure the same physical quantity $F$.
\footnote{Of course here we are considering two observers: $O$ and an inertial one, each of them in a different space-time where the same physical entity is present, for instance an electric field.}

We shall say that the Principle of General relativity (GRP) is verified when the following two conditions hold:
\begin{description}
\item{(a)} The equations describing the behaviour of the physical quantities have the same form in all sufficiently regular coordinate systems with the metric ${\bf g}$ being the only quantity pertaining to space-time that can appear in those equations; moreover, the latter become on $C$ their corresponding ones in Minkowskian coordinates in flat space-time $({\bf M}, {\bf \eta})$  when they are expressed in terms of the $\tilde{x}^{\lambda}$'s if $\mathbf{a}=\mathbf{\omega}=0$. (This implies the Equivalence Principle in general, but curvature dependent terms may still appear if there is no minimal coupling or higher order derivatives are involved in the equations.)
\item{(b)} One has $\tilde{Q}[\tilde{F}]\subseteq Q[F],~\forall ~F$ and $~\forall~$QMC$C$$\mathbf{\omega}$.  
\end{description}

Some comments are in order:
\begin{description}
\item{(i)} Note that if $\tilde{x}^{\lambda}$ and $\tilde{x}'^{\lambda}$ are two different systems of QMC$C\mathbf{\omega}$'s for $O$ -with common $\mathbf{a}, \mathbf{\omega}$ and $\mathbf{e}_{(\alpha)}$-~in $({\bf M}, {\bf g})$, one has in general that $\tilde{Q}[\tilde{x}^{\lambda}] \neq \tilde{Q}'[\tilde{x}'^{\lambda}]$, and if 	$x^{\lambda}$ are Minkowskian coordinates in flat space-time $({\bf M}, {\bf \eta})$ and $q[x^{\lambda}] \in Q[x^{\lambda}]$, there will be QMC$C\mathbf{\omega}$'s, $\tilde{x}^{\lambda}, \forall O(C,\mathbf{\omega})$, such that $q[x^{\lambda}] \in \tilde{Q}[\tilde{x}^{\lambda}] \subseteq Q[x^{\lambda}]$.
\item{(ii)}Let us suppose that ${\bf g}$ admits an isometry  $\Lambda$ and let the isometric transformation induced on coordinates and physical quantities be $\Lambda:x
\longrightarrow x'$, $\Lambda:F(P,x)\longrightarrow F'(P,x')$. Then if one accepts that, as we shall show in Section IV, \emph{it is the functional form of the metric that determines the physical meaning of its coordinate arguments}, condition (a) of the GRP implies $Q[F(P,x)]=Q[F'(P,x')]$ since the mathematical expressions of all physical laws involving $F(P,x)$ and $F'(P,x')$ are formulated in terms of coordinates $x$ and $x'$ having the same physical meaning and have exactly the same form. Thus a GRP implies a RRP with respect to any existing isometry.
\item{(iii)}Condition (a) of the GRP entails the diffeomorphism invariance but this just by itself has not much physical significance as was pointed out by Kretsmann~\cite{kr} in 1917 and illustrated by Cartan~\cite{ca} in 1923 and Friedrichs~\cite{fr} in 1927 with their diffeomorphism invariant formulations of Newtonian gravitation. It is the rest of condition (a) and condition (b) what really confers physical meaning to the stated GRP. It might be objected that the latter requirements are expressed in terms of the special family of coordinates QMC$C$$\mathbf{\omega}$, $\tilde{x}^{\lambda}$'s. But that has been done just for the sake of simplicity in our formulation of the GRP. In fact, let us consider coordinates QMC$C$$\mathbf{\omega}$, $\tilde{x}^{\lambda}$'s, in a generic space-time and Minkowskian coordinates in flat space-time, $x^{\lambda}$'s, and the two formally identical transformations: $\phi: \tilde{x}\to \tilde{x}', \tilde{x}'^{\lambda}= \phi^{\lambda}(\tilde{x})$ and  $\phi: x\to x', x'^{\lambda}= \phi^{\lambda}(x)$; then if the mathematical expressions of the physical laws, when written in terms of the $\tilde{x}^{\lambda}$'s, have the same form on $C$ than their corresponding ones in Minkowskian coordinates, $x^{\lambda}$'s, in flat space-time, they still will have the same forms when respectively written in terms of the coordinates $\tilde{x}'$ on $C$ and $x'$. Furthermore, it also follows that $\tilde{Q}[\tilde{x}'^{\lambda}] \subseteq Q[x'^{\lambda}]$ and $\tilde{Q}[\tilde{F}']\subseteq Q[F']$, for any physical quantity in a generic space-time, $\tilde{F}'$, and flat space-time, $F'$, respectively expressed in terms of the coordinates $\tilde{x}'$ and $x'$. Thus, although a special type of coordinates has been used in the above definition of the GRP, it has a fundamental significance that is independent of the class of coordinate systems considered. 
\end{description}
\subsubsection{ Principle of General Covariance (GCP)}

The GCP is the above GRP. 
\begin{itemize}
\item[-]It implies the Equivalence Principle.
\item[-] The physical content of the GCP is thus that one given in the previous definition of the GRP.   
\end{itemize}
What is called for now is to verify that this interpretation of the GCP is coherent with the theory and agrees with experience. That is therefore tantamount to testing the validity of the stated GRP. To that end we proceed to construct the QMC$C$$\mathbf{\omega}'s$ in the next Section.

\section{Construction of the QMC$C$$\mathbf{\omega}$'s}\label{sec:QMC}
The $\tilde{x}^{\lambda}$'s will be constructed under the assumption that the $x^{\lambda}$'s can be expressed  as power series of the $\tilde{x}^{i}$'s with coefficients depending on $\tau$ about $C$.  The transportation law (2) sets the following constraints on $C$ upon the Christoffel symbols $\tilde{\Gamma}_{\mu \nu}^{\rho}(\tau)$ and $\tilde{\Gamma}_{\mu \nu \rho}(\tau)$:

\begin{equation}
\tilde{\Gamma}_{00}^{0}=\tilde{\Gamma}_{000}=0, \tilde{\Gamma}_{k0}^{0}=-\tilde{\Gamma}_{0k0}=\tilde{\Gamma}_{k00}=\tilde{\Gamma}_{00}^{k}=\frac{1}{c^{2}}\tilde{a}^{k}, 
\tilde{\Gamma}_{k0}^{j}=\tilde{\Gamma}_{jk0}=-\frac{1}{c}\tilde{\omega}^{i}\tilde{\epsilon}_{0ijk},
\end{equation}
where the tildes always indicate that the components of the quantities are in the QMC$C$$\mathbf{\omega}$ coordinates.

Noting that $\tilde{a}^{0}=\tilde{\omega}^{0}=0$ and assuming that the
remaining Christoffel symbols on $C$ may be expressed as power series of
the components of the 4-acceleration and rotation, condition 2  of
subsection \ref{ssec:GRP} allows us to put for them
\begin{equation}
\tilde{\Gamma}_{jk}^{\alpha}(\tau)=p_{jki}^{\alpha}(\tau)\tilde{a}^{i}+q_{jki}^{\alpha}(\tau)\tilde{\omega}^{i}+O_{jk}^{\alpha}(2),
\end{equation}
 with $p_{jki}^{\alpha}(\tau)$ and $q_{jki}^{\alpha}(\tau)$ being smooth enough but otherwise arbitrary functions of the proper time along $C, \tau$, and $O_{jk}^{\alpha}(2)$ similarly standing for a second order term in the $\tilde{a}^{i}$'s and $\tilde{\omega}^{i}$'s. Obviously all these functions are symmetric in the indices $j$ and $k$. Eq. (4) may be rewritten in terms of the components of $\mathbf{a}$ and $\mathbf{\omega}$ in  the given coordinates $x$ as 
 
 \begin{equation}
\tilde{\Gamma}_{ik}^{\alpha}(\tau)=A_{ik\nu}^{\alpha}(\tau)a^{\nu}+B_{ik\nu}^{\alpha}(\tau)\omega^{\nu}+O_{ik}^{\alpha}(2)~,\end{equation}
where the  $A_{ik\nu}^{\alpha}$ and $B_{ik\nu}^{\alpha}$ are arbitrary save by being sufficiently smooth and the constraints 

\begin{equation}
A_{ik\nu}^{\alpha}=A_{ki\nu}^{\alpha}~,~B_{ik\nu}^{\alpha}=B_{ki\nu}^{\alpha}~,~A_{ik\nu}^{\alpha}e^{\nu}_{(0)}=B_{ik\nu}^{\alpha}e^{\nu}_{(0)}=0~;\end{equation}
 and $O_{ik}^{\alpha}(2)=O_{ki}^{\alpha}(2)$ is of second order in the $a^{\nu}$'s and $\omega^{\nu}$'s, otherwise $O_{ik}^{\alpha}(2)$ is a function of $\tau$ as so is $\Gamma_{\rho\sigma}^{\lambda}(\tau)$, as both are defined on $C$.

Taking into account condition 3 of subsection \ref{ssec:GRP} and using the transformation law for the metric connection on $C$ we get  
\begin{equation}\label{eq:7}
x^{\lambda}=f^{\lambda}(\tau)+e^{\lambda}_{(k)}\tilde{x}^{k}+\frac{1}{2}\big(e^{\lambda}_{(\alpha)}\tilde{\Gamma}_{ik}^{\alpha}(\tau)-e^{\rho}_{(i)}e^{\sigma}_{(k)}\Gamma_{\rho\sigma}^{\lambda}(\tau)   \big)\tilde{x}^{i}\tilde{x}^{k}+\Phi^{\lambda}(\tilde{x})~,
\end{equation}
with $\Phi^{\lambda}(\tilde{x})$ being of third order in the $\tilde{x}^{i}$'s. In order to ensure the fulfilment of condition 4 of subsection \ref{ssec:GRP} we split $\Phi^{\lambda}(\tilde{x})$ into two terms as follows:
\begin{equation}
\Phi^{\lambda}(\tilde{x})=\Psi^{\lambda}(\tilde{x})+\Phi^{\lambda}_{(0)}(\tilde{x})~,\end{equation}
with $\Psi^{\lambda}(\tilde{x})$ verifying
\begin{equation}
\Psi^{\lambda}\mid_{C}= \frac{\partial\Psi^{\lambda}}{\partial\tilde{x}^{i}}\mid_{C}=\frac{\partial^{2}\Psi^{\lambda}}{\partial\tilde{x}^{i}\partial\tilde{x}^{j}}\mid_{C}=0~,\end{equation}
and vanishing whenever ${\bf a}, {\bf \omega}$ and the curvature tensor, ${\bf R}$, in a finite neighborhood of $C$, all vanish; the latter is equivalent to the vanishing of the $\tilde{\Gamma}_{\mu\nu}^{\lambda}$'s in that neighborhood; otherwise the $\Psi^{\lambda}$'s apart from being sufficiently smooth are arbitrary; it is  $\Phi^{\lambda}_{(0)}(\tilde{x})$ that should be  determined to assure that the coordinates $\tilde{x}$ have the property specified in condition 4 of subsection \ref{ssec:GRP}. To that end we put  
\begin{equation}
\Phi^{\lambda}_{(0)}(\tilde{x})=\sum_{l,m,n}\frac{1}{l!m!n!}C_{lmn}^{\lambda}(\tau)(\tilde{x}^{1})^{l}(\tilde{x}^{2})^{m}(\tilde{x}^{3})^{n}~,\end{equation}
 with \begin{displaymath}l\ge0,~ m\ge0,~ n\ge0,~ l+m+n\ge3,~\textrm{and}~ l,m,n ~\textrm{all being integers},\end{displaymath}
and the $C_{lmn}^{\lambda}(\tau)$'s are systematically calculated by the following algorithm:
Consider the equation
\begin{equation}
\frac{\partial^{2}\Phi^{\lambda}_{(0)}}{\partial\tilde{x}^{i}\partial\tilde{x}^{k}}=-\frac{\partial x^{\rho}}{\partial\tilde{x}^{i}}\frac{\partial x^{\sigma}}{\partial\tilde{x}^{k}} \Gamma_{\rho\sigma}^{\lambda}(x(\tilde{x}^{\mu}))+e^{\beta}_{(i)}e^{\gamma}_{(k)}
\Gamma_{\beta\gamma}^{\lambda}(\tau)~,\end{equation}
where the first term on the rhs is taken as dependent, in general, on the $\tilde{x}^{\mu}$'s, while the second term only depends on $\tau$ as is evaluated on $C$. Eq. $(11)$ is a consequence of considering the equation for the transformation of the Christoffel symbols on $C$ and using eqs. $(7)$ and $(8)$ besides setting $\tilde{\Gamma}_{\mu\nu}^{\alpha}=\Psi^{\lambda}=0$. The $C_{lmn}^{\lambda}(\tau)$'s are found by using the power series for $\Phi^{\lambda}_{(0)}$ given in eq. $(10)$ and taking successive derivatives of eq. $(11)$ with respect to the $\tilde{x}^{j}$'s, taking the result on $C$, and doing it all along as if $\tilde{\Gamma}_{\mu\nu}^{\alpha}=\Psi^{\lambda}=0$  at all points.

So we get, for instance,
\begin{displaymath}
C_{300}^{\lambda}(\tau)=-\frac{\partial}{\partial\tilde{x}^{1}}\bigg(\frac{\partial x^{\rho}}{\partial\tilde{x}^{1}}\frac{\partial x^{\sigma}}{\partial\tilde{x}^{1}}\Gamma_{\rho\sigma}^{\lambda}(x)\bigg)\arrowvert_C=-4e^{\sigma}_{(1)}A^{\rho}_{11}\Gamma_{\rho\sigma}^{\lambda}-e^{\rho}_{(1)}e^{\sigma}_{(1)}\Gamma_{\rho\sigma,\gamma}^{\lambda}e^{\gamma}_{(1)},\end{displaymath} with $A^{\lambda}_{ik}=-\frac{1}{2}e^{\rho}_{(i)}e^{\sigma}_{(k)}\Gamma_{\rho\sigma}^{\lambda}(\tau)$, all evaluated on $C$ at the point corresponding to $\tau$.
That way any $C_{lmn}^{\lambda}(\tau)$ may be expressed in terms of the $C_{l'm'n'}^{\lambda}$'s of lower order: $l'+m'+n'<l+m+n,~l'\le l,~m'\le m,~n'\le n$;~the $e^{\lambda}_{(k)}(\tau)$, the $\Gamma_{\mu\nu}^{\alpha}(\tau)$'s, and the partial derivatives of the $\Gamma_{\mu\nu}^{\alpha}$'s with respect to the $x^{\lambda}$'s up to order $\le l+m+n-2$.

Fixing the $\Psi^{\lambda}$'s, $A_{ik\nu}^{\alpha}$'s,  $B_{ik\nu}^{\alpha}$'s and the $O_{ik}^{\alpha}(2)$'s uniquely determines a set of corresponding QMC$C$$\mathbf{\omega}$. If space-time is flat and also $\mathbf{a}=\mathbf{\omega}=0$, it follows that $\Psi^{\lambda}= \tilde{\Gamma}_{\mu\nu}^{\alpha}(\tilde{x})=0$, and the entire family of the QMC$C$$\mathbf{\omega}$'s collapses to the unique usual Lorentzian coordinates corresponding to the chosen tetrad $e^{\lambda}_{(\nu)}$. This does not mean that if space-time is not flat and/or if $\mathbf{a}\not =0$, or $\mathbf{\omega}\not =0$, by taking $\Psi^{\lambda}=A_{ik\nu}^{\alpha}=B_{ik\nu}^{\alpha}=O_{ik}^{\alpha}(2)=0$, the resulting QMC$C$$\mathbf{\omega}$ would be Lorentzian, as these do not symply exist for non-flat space-times or in non-inertial reference frames.

\section{The Hole Argument}\label{sec:hole}
The Hole Argument (HA) is a consideration that was first raised by
Einstein in a letter of November 2, 1913, to Ludwig Hopf. He was then
struggling to find the gravitational field equations and intended to
prove with the HA, whose final Einsteinian version was published in 1914
~\cite{ein}, that the theory could not be generally covariant. Obviously
Einstein had discarded that implication of the HA by November, 1915,
with his settling upon the correct generally covariant field equations.
The HA has been widely discussed in the literature (see, for instance,
Stachel~\cite{st}(1980, 2002), Rovelli~\cite{ro} (1991, 2008),
Norton~\cite{no} (1993) and Lusanna and Pauri~\cite{lu} (2006)). Here we
present a treatment of the HA by resorting to the idea of the
QMC$C$$\mathbf{\omega}$'s introduced in subsection \ref{ssec:GRP} that we believe
further clarifies the whole issue, specially by showing how it is the
metric field that supplies the physical meaning of coordinates and
individuates point-events in regions of space-time where no other fields
exist. 

Essentially the HA goes as follows: Let us assume there is an  region $\mathnormal{H}$ (the hole) where all the nongravitational fields are null. Let that region be covered by coordinates $x^{\lambda}$ that extend to a larger coordinate patch $U$: $\mathnormal{H} \subset U$. Let us consider a coordinate transformation 
\begin{equation}\label{eq:12}
\phi : x \to x', x'^{\lambda}= \phi^{\lambda}(x), 
\end{equation}
that smoothly becomes the identity transformation outside $H$ and on its boundary. Under the transformation the components of the metric tensor change according to 
\begin{equation}
g_{\mu\nu}(x) \to g'_{\mu\nu}(x') = \frac{\partial x^{\alpha}}{\partial x'^{\mu}}\frac{\partial x^{\beta}}{\partial x'^{\nu}}g_{\alpha\beta}(x(x'))
\end{equation}
Let $\sigma$ be the map of $U$ into $\Re^{4}$ that assigns the $x$ coordinates to the point events in $U$ and let us suppose that $\phi\circ\sigma(H) =  \sigma(H)\subset\Re^{4}$ and also that all the coordinates involved, $x$ as well as $x'$, are homogeneous quantities, say everyone of them is a length value. Then the GC of the Einstein equations assures that the metric $g'_{\mu\nu}(x)~ \forall ~x \in \sigma(U)$ provides a new solution to those equations if the argument coordinates in $g'_{\mu\nu}$ and $g_{\mu\nu}$ are interpreted as designating the same events whenever they take the same values. This is the HA that leads to different solutions for a single mass-energy distribution and hence to a supposed inadequacy of GR as a consequence of its GC. It would violate causality in an obvious sense. 

We shall see now how the introduction of QMC$C$$\mathbf{\omega}$'s dispels the difficulty posed by the HA. Let us have a set of QMC$C$$\mathbf{\omega}$'s, $\tilde{x}$, covering $\mathnormal{H}$ or part of it. Let the given coordinates, $x$, be related to the $\tilde{x}$ by
\begin{equation}\label{eq:14}
x^{\lambda}=\chi^{\lambda}(\tilde{x})
\end{equation}
that under suitable conditions might be expressed as in eq. $(\ref{eq:7})$ A concrete event in $H$ may be labelled by its coordinates $\tilde{x}$ as these are uniquely assigned to that event by a specific operational protocol. The particular physical process that the measurement act entails individuates the corresponding event. 
Eqs. (\ref{eq:12}) and (\ref{eq:14}) give
\begin{equation}
x'^{\lambda}= \phi^{\lambda}(x)= \phi^{\lambda}(\chi(\tilde{x})) \equiv \rho^{\lambda}(\tilde{x})
\end{equation}
In the preceding exposition of the HA the $x'$ in the functions $g'_{\mu\nu}$ were called $x$ to get the \emph{new} metric $g'_{\mu\nu}(x)$. Coherently with that we should rewrite the above equation as 
\begin{equation}
x^{\lambda}=  \rho^{\lambda}(\tilde{x})
\end{equation}
and since the functions $\chi^{\lambda}$ and $\rho^{\lambda}$ are different on $H$, unless $(12)$ is the identity transformation, one may not have in general the same values for the $\tilde{x}^{\alpha}$ on the rhs's of eqs. $(14)$ and $(16)$ if one insists on having identical values for the $x^{\lambda}$ on the lhs's of those equations. Thus we will have in general 
\begin{equation}
x^{\lambda}=\chi^{\lambda}(\tilde{x}_{1})=\rho^{\lambda}(\tilde{x}_{2}),  
\end{equation}
with $\tilde{x}_{1} \neq \tilde{x}_{2},$
indicating that we are dealing in general with distinct events when they are labeled by the same values in the coordinates $x$ and $x'$. Since the $x^{\lambda}$ and the $x'^{\lambda}$ are different functions of the $\tilde{x}^{\alpha}$ which have been given a precise operational meaning, the $x^{\lambda}$ and the $x'^{\lambda}$ have a distinct physical interpretation. The relationships of $x$ and $x'$ to $\tilde{x}$ clearly respectively depend on the functional forms of the metric tensor $g_{\mu\nu}(x)$ and $g'_{\mu\nu}(x')$ in terms of those coordinates as explicitly follows from our construction in eq. $(7)$ and implicitly and more generally from eqs. $(12), (13), (14)$ and $(15)$. Therefore one has to conclude that \emph{it is the functional form of the metric that determines the physical meaning of its coordinate arguments}. Thus in GR coordinates in space-time are physically meaningless before specifying the metric tensor though they designate a particular point of the underlying mathematical manifold ${\bf M}$, as has been pointed out differently by Stachel~\cite{st} and Norton~\cite{nor}(2002). This way one clearly sees that the HA is no objection to the requirement of GC for a metric theory such as GR. The preceding conclusion has an interesting corollary:  \emph{Let us ask ourselves if it would be possible, in a metric theory such as GR, to have two} \emph{\textbf {different}} \emph{space-times of respective metrics $g_{\mu\nu}(x)$ and $g'_{\mu\nu}(x')$ functionally related by eq. $(13)$ and such that the physical (operational) meanings of the coordinates $x$ and $x'$ were the same. The answer should be in the negative!}

There is an alternative coordinate-independent way of presenting the HA that was first pointed out by Stachel in 1980~\cite{st}. It is essentially equivalent to the one just given, but it has customarily  become the modern account  of the HA and provides other insights regarding the conclusions reached at the end of the preceding paragraph, particularly concerning the physical individuation of point-events in space-time as a consequence of the metric field. We will sketch it here for completeness and to find that both descriptions complement each other illuminating part of the deep significance of the GCP.

Let $\phi: M \to M$ be a sufficietly differentiable diffeomorphic map that becomes the identity map outside $H$ and on its boundary so that also $\phi: H \to H$. Let $p$ be an arbitrary point belonging to $H$, and $x$  its given coordinates in a certain chart containing $p$. Likewise let $x'$ be the coordinates of the diffeomorphic image of $p, \phi(p),$ in another chart covering this last point that may coincide or not with the former chart. We shall also denote by eq. (12) the functional correspondence induced by the diffeomorphism between the coordinates in the two charts associated to some neighbourhoods of $p$ and $\phi(p)$. It is well known that the \emph{active} diffeomorphism $\phi$ also generates a drag-along $\phi^{\ast}$ from tensors at $p$ to $\phi(p)$. In particular the drag-along metric tensor at $\phi(p), \phi^{\ast}{\bf g}$, has components $g'_{\mu\nu}(x')$ in the $x'$ coordinates verifying eq. (13) above, with the terms $g_{\alpha\beta}(x)$ entering its rhs being now the components of ${\bf g}$ at $p$ in the $x$ coordinates. The GC of GR again implies that the \emph{new} metric $\phi^{\ast}{\bf g}$ satisfies as well the Einstein equations. It is true that the tensors ${\bf g}$ and $\phi^{\ast}{\bf g}$ would be the same were they attached at the same point and the different coordinates $x$ and $x'$ corresponded to that point, but that is not the case. One has now a relocation of the metric field over the points of $H$ in which  $\phi^{\ast}{\bf g}$ is at $\phi(p)$, whereas ${\bf g}$, its geometrical equivalent, was at $p$ before the drag-along. The answer to this version of the HA has been to assert that a unique physical solution of the Einsten equations is given by the class of equivalence, $\{(M, \phi^{\ast}{\bf g}), \forall \phi\}$, obtained by considering the action of all possible diffeomorphisms of the previous kind in the way just explained. That equivalence has been called \emph{Leibniz equivalence} in the literature. Along with this emerges the idea that the point-events of space-time are only individuated by the physical entities present at them, in our case by only the metric field that is the only physical field existing in the hole. In the language of the QMC$C$$\mathbf{\omega}$ coordinates, $\tilde{x}$: \emph{the two manifold points $p$ and $\phi(p)$, with the respectively attached metrics, ${\bf g}$ and $\phi^{\ast}{\bf g}$, correspond to the same event in space-time as their coordinates, $x^{\lambda}$ and $x'^{\lambda}$, are related to the same values of the individuating $\tilde{x}'s$ by the functions $\chi^{\lambda}(\tilde{x})$ and $\rho^{\lambda}(\tilde{x})$ whose forms -as was pointed out above- are precisely determined by the components of the mentioned two geometrically equivalent metrics.}

\section{Remarks and conclusions}\label{sec:remarks}
Consider a specific but generic observer $O$ of world-line $C$ who uses a chosen set of QMC$C$$\mathbf{\omega}$, $\tilde{x}^{\lambda}$'s. Let $P$ be the space-time position of $O$ at its proper time $\tau_{P}$ and let $\delta x^{\alpha}$ the components of an infinitesimal 4-vector with origin at $P$ in the given coordinates $x^{\lambda}$. One may put 
$$\delta x^{\alpha}=\delta x^{\alpha}_{\parallel}+\delta x^{\alpha}_{\perp}~,$$ where the last two terms respectively stand for the the parallel and perpedicular parts of $\delta x^{\alpha}$ to $\mathbf{e}_{0}$. The proyector on the hyperplane normal to $\mathbf{e}_{0}$ is
$$g_{\alpha\beta}+\frac{1}{c^{2}}u_{\alpha}u_{\beta}~.$$ 
Then the quantity 
\begin{equation}
dl^{2} \equiv ( g_{\alpha\beta}+\frac{1}{c^{2}}u_{\alpha}u_{\beta})\delta x^{\alpha}\delta x^{\beta}= \delta x^{\alpha}_{\perp}\delta x_{{\alpha}_{\perp}}~,
\end{equation}
 certainly is $\delta\tilde{x}^{i}\delta\tilde{x}_{i}$ and should therefore be interpreted as the spatial distance squared measured by $O$ at time $\tau_{P}$ between $P$ and the point-event $Q$ at the tip of the vector $\delta x^{\alpha}$. This is consistent with what follows from inverting eq. (7) neglecting higher order terms in the infinitesimals.

If the observer $O$ happens to be in free fall, momentarily at rest at $\tau_{P}$ and his reference tetrad does not rotate he would then measure the proper distance between $P$ and $Q$, and eq. (18) above yields for the square of that   
\begin{equation}
dl^{2}=\delta_{ik}\delta\tilde{x}^{i}\delta\tilde{x}^{k} = ( g_{\alpha\beta}-\frac{g_{0\alpha}g_{0\beta}}{g_{00}} ) \delta x^{\alpha}\delta x^{\beta}~,
\end{equation}
that is the usually accepted result~\cite{mo}.

The relationships of the generic given coordinates $x^{\lambda}$ and two different sets of QMC$C$$\mathbf{\omega}$, $\tilde{x}^{\lambda}$ and $\tilde{x}'^{\lambda}$, may differ at most by terms of second order in the $\tilde{x}^{i}$'s and $\tilde{x}'^{i}$'s if the observer is not in free fall ($C$ is not a geodesic) and/or his/her choiced transported tetrad rotates, which corresponds to the freedom allowed to choose the $\tilde{\Gamma}_{ik}^{\alpha}(\tau)$'s via eq. $(5)$, or by terms of third order in the same variables if the observer is in free fall and its reference tetrad is paralell transported, corresponding to the freedom allowed to choose the function $\Phi^{\lambda}$ when the space-time is not flat. 

It follows from eq. $(7)$ that the values of tensor quantities measured on the world-line $C$ corresponding to two different sets of QMC$C$$\mathbf{\omega}$'s, $\tilde{x}$ and $\tilde{x}'$, -but with the same choice of reference tetrad- should be identical. However if these quantities are measured by the observer at small $\delta\tilde{x}^{k}$, equivalently $\delta\tilde{x}'^{k}$, off his world-line $C$ one has, for instance and with no loss of generality, for the components of the electromagnetic field tensor, $F^{\lambda\mu}$, in the two sets of coordinates: 
$$
\tilde{F}^{\lambda\mu}=\tilde{F}'^{\lambda\mu}+(C_{ik\nu}^{\lambda}\tilde{F}'^{i\mu}
+C_{ik\nu}^{\mu}\tilde{F}'^{\lambda i})a^{\nu}\delta\tilde{x}'^{k}+(D_{ik\nu}^{\lambda}\tilde{F}'^{i\mu}+D_{ik\nu}^{\mu}\tilde{F}'^{\lambda i})\omega^{\nu}\delta\tilde{x}'^{k}+ h.o.t.,
$$
where eqs. $(7)$ and $(5)$ have been used and we have put
$C_{ik\nu}^{\lambda} \equiv A_{ik\nu}'^{\lambda}-A_{ik\nu}^{\lambda}$
and $D_{ik\nu}^{\lambda} \equiv
B_{ik\nu}'^{\lambda}-B_{ik\nu}^{\lambda}$, $A_{ik\nu}'^{\lambda}$ and
$B_{ik\nu}'^{\lambda}$ being the coefficients that correspond to
${\tilde{\Gamma}}_{ik}'^{\alpha}(\tau)$ in its expression analogous to eq.
$(5)$ for $\tilde{\Gamma}_{ik}^{\alpha}(\tau)$; $h.o.t.$ stands for a
series of first or higher  order terms in the spatial displacements
$\delta\tilde{x}'^{k}$ and higher order terms in the $a^{\nu}$ and/or
$\omega^{\nu}$ when the $\delta\tilde{x}'^{k}$ occur only to first
order. That means that in general, in the vicinity of the observer, the
discrepancies between the outcomes of the different \emph{standard}
measurement methods of the same physical quantity are more sensitive to
local inertial effects, when they exist, than to gravitational fields.

After all the foregoing considerations we can say that the meaning of
the GCP is, at least, two-fold: On the one side, as a GRP such as we
defined it in Section \ref{sec:GCP}, it is a really predictive physical
principle like the SRP, but with the generalizations and conditions
specified thereby, namely replacement of inertial observers by general
ones, of Minkowskian coordinates by quasi- Minkowskian ones, the
appearance of the space-time metric as a new physical tensor quantity
and the splitting induced in the results of the measurements of the same
physical quantities when different measurement protocols are used though
they be equivalent in the absence of gravitation or inertial effects. On
the other side, as has been shown in Section \ref{sec:hole} with our discussion of
the HA, it provides deep insights on how the nature of coordinates
depends on the form of the gravitational fields that, consistently with
that, are the entities that individuate -or, together with other
physical entities that might be present, contribute to individuate- the
point-events of space-time. Because of that the GCP has also value as a
guiding principle supporting Einstein's appreciation of its heuristic
worth in his reply to Kretschmann~\cite{ei} (1918). So it would seem to
favor quantum theories of gravitation without \emph{a priori} background
space-time againts those theories that assume such background structure
\emph{ab initio} as has also been pointed out by Lusanna and
Pauri~\cite{lu} (2006).

\section*{Acnowledgments}
I would like to thank Professor Jes\'us Mart\'in for useful and detailed conversations and Professor Llu\'is Bel for his always interesting comments and suggestions. This work was started long ago while I was on sabbatical leave at the Department of Atomic, Molecular and Nuclear Physics of the University of Seville by the kind invitation of Professor Lu\'is Rull.


\begin{thebibliography}{99}
\bibitem{kr} E.~Krestchmann, \emph{Annalen der Physik}, {\bf 53}, 575 (1917).
\bibitem{ei} A.~Einstein, \emph{Annalen der Physik}, {\bf 55}, 240 (1918).
\bibitem{ca} E.~Cartan, \emph{Ann. Sci. ENS}, {\bf40}, 325 (1923).
\bibitem{fr}K.~Friedrichs,\emph{Matematische Annalen}, {\bf98},566 (1927).
\bibitem{mi}C.W.Misner, K.S.Thorne and J.A.Wheeler,\emph{Gravitation}(Freeman, San Francisco, 1973).
\bibitem{fo}V.~Fock,\emph{The Theory of Space, Time and Gravitation}(Pergamon Press, 2nd Revised Ed. 1966), pp. 5-8, 178-182 and 392-396.
\bibitem{an}J.L.~Anderson, \emph{Principles of Relativity Physics}(Academic Press, New York,1967).

\bibitem{st}J.~Stachel, ``Einstein's Search for General Covariance,
1912-1915'', in \emph{Einstein from `B' to `Z'}(Birkh\"{a}user,Boston,
2002), pp. 301-337. [This paper was first read at the Ninth
International Conference on General Relativity and Gravitation, Jena,
Germany, 1980];

 ``What a Physicist Can Learn from the History of Einstein's Discovery
 of General relativity'', in \emph{Proceedings of the Fourth Marcel
 Grossmann Meeting on General Relativity}, R. Ruffini, ed.
 (Elsevier,Amsterdam, 1986), pp. 1857-1862. 
\bibitem{no}J.D.~Norton,\emph{Rep. Prog. Phys.}, {\bf56}, 791 (1993).
\bibitem{el}G.F.R.Ellis and D. R.Matravers, \emph{Gen. Rel. Grav.}, {\bf27}, 777 (1995).
\bibitem{wa} R.M.~Wald, \emph{General Relativity} (The University of Chicago Press, Chicago, 1984), pp. 58-60.
\bibitem{la} M.~Lachi\`{e}ze-Rey, gr-qc/ 0107010 v2 (2001).
\bibitem{lu}L.~Lusanna and M.~Pauri, \emph{Gen. Rel. Grav.}, {\bf38(2)}, 187 and 229 (2006).

\bibitem{de}See an illustration of that splitting of operational
prescriptions for determining QMC's in uniformly accelerated reference
frames in E.A.~Desloge and R.J.~Philpott, \emph{Am. J. Phys.}, {\bf55},
252 (1987). 

\bibitem{ein} A.~Einstein, \emph{``Die formale Grundlage der allgemeinen Relativit\"atstheorie,'' K\"oniglich Preusssche Akademie de Wissenschaften (Berlin), Sitzungsberichte}, p. 1067 (1914).
\bibitem{ro} C.~Rovelli, \emph{Class. Quantum Grav.}, {\bf8}, 297 (1991); \emph{Quantum Gravity} (Cambridge University Press, Paperback Ed. 2008), pp. 65-71.
\bibitem{nor}J.D.~Norton, ``Einstein's Triumph over the Space-Time Coordinate System'', in \emph{Di\'{a}logos} {\bf79}, pp. 253-262 (2002).
\bibitem{mo}See, for instance, C.~M{\o}ller, \emph{The Theory of Relativity} (Oxford University Press, 2nd Ed. 1972).
\end{thebibliography}
\end{document}